\documentclass[twocolumn]{article}%
\usepackage{amsmath}
\usepackage{amsfonts}
\usepackage{amssymb}
\usepackage{graphicx}%
\providecommand{\U}[1]{\protect\rule{.1in}{.1in}}
\newtheorem{theorem}{Theorem}

\newtheorem{remark}[theorem]{Remark}

\begin{document}

\title{Superluminal Neutrinos from OPERA Experiment and Weyl Equation}
\author{E. Capelas de Oliveira\thanks{capelas@ime.unicamp.br}, W. A. Rodrigues
Jr.\thanks{walrod@ime.unicamp.br\ or walrod@mpc.com.br}. and J. Vaz
Jr.\thanks{vaz@ime.unicamp.br}\\$\hspace{-0.1cm}$IMECC-UNICAMP\\University of Campinas\\13083-859 Campinas, SP, Brazil}
\maketitle

\begin{abstract}
By analyzing the structure of the Weyl spinor field in the Clifford bundle
formalism we show that in each spinorial frame it is represented by
$\mathcal{F\in}\sec(%
{\textstyle\bigwedge\nolimits^{0}}
T^{\ast}M+%
{\textstyle\bigwedge\nolimits^{2}}
T^{\ast}M+%
{\textstyle\bigwedge\nolimits^{4}}
T^{\ast}M)\hookrightarrow\sec\mathcal{C\ell(}M,g)$ satisfying the equation
$\boldsymbol{\partial}\mathcal{F}=0$, where $\boldsymbol{\partial}$ is the
Dirac operator acting on sections of the Clifford bundle $\mathcal{C\ell
(}M,g)$. With this result we show that introducing a generalized potential
$\mathcal{A}=(A+\gamma_{5}B)\mathcal{\in}\sec(%
{\textstyle\bigwedge\nolimits^{1}}
T^{\ast}M+%
{\textstyle\bigwedge\nolimits^{3}}
T^{\ast}M)\hookrightarrow\sec\mathcal{C}\ell(M,g)$ for the Weyl field such
that \ $\mathcal{F}=\boldsymbol{\partial}\mathcal{A}$ it is possible to
exhibit superluminal solutions (including one with a \emph{front }moving at
superluminal speed) for Weyl equation, which\ surprisingly describes the
propagation of a \textit{massive} tachyonic neutrino. We propose to interpret
these extraordinary solutions in order that eventually\ they may serve as
possible models for the emission process and propagation of the superluminal
neutrinos observed at the OPERA experiment. Moreover, complementing this study
we show that general local chiral invariance of Weyl equation implies that it
describes for all solutions that are eigenstates of the parity operator a pair
of `sub-particles' carrying opposite magnetic charges (thus possibly carrying
a small magnetic moment) which thus interact with an external electromagnetic
field. Even if at the Earth's electromagnetic field the effect may result
negligible, eventually the idea may be a useful one to study neutrinos leaving
the electromagnetic field of stars.

\end{abstract}

\section{Introduction}

In this paper we show that Weyl equation that was originally thought to
describe the propagation at the speed of light of massless neutrinos has also
some nontrivial \emph{subluminal} and \emph{superluminal} solutions that
simulate the propagation of a particle with non zero rest mass, respectively
at subluminal and superluminal speeds. We even exhibit a superluminal solution
of Weyl equation \ occupying a compact support in the direction of propagation
for any instant of time, where its front moves at superluminal speed. In order
to exhibit such solutions we shall need to show that the Weyl field\ when
represented in the Clifford bundle formalism (used in this paper) posses a
superpotential that satisfies the homogeneous wave equation. To prove this
statement we need to recall some mathematical results\footnote{A detailed
presentation of the Clifford bundle formalism may be found in
\cite{rodcap2007}.}, unfortunately not so well known by physicists. This is
done in Section 2. Eventually our results will serve the purpose of giving a
possible explanation to the OPERA experiment\footnote{A bunch of
papers\ `explaining' the OPERA experiment appeared in the arXiv in the last
week, e.g., \cite{iorio,Keh,kono,oiko,tamblav}. We are not going to comment on
all them here.} \cite{adam} and may even explain the discrepant results
concerning the square of the neutrino\ `mass' in different experiments done
under different circumstances
\cite{assa,bionta,cab,ciboremb1996,ciboremb1998,ciboremb2011,hira,longo}. The
authors of the present paper are well aware of some criticisms to the OPERA
experiment\footnote{We want also to state that we are well aware about the
many claims concerning superluminal propagation of microwaves (and even single
photons) \cite{enders1993,mug,stei1993} and which has been the subject of many
misunderstandings, since all those phenomena has a common simply explanation,
namely \emph{pulse reshaping} \cite{rtx2001,crtx2001} and thus do not implies
in any breakdown of the Principle of Relativity. Thus, authors do not exclude
that results of OPERA experiment will eventually find an explanation which
does not implies in any violation of the Principle of Relativity.} that range,
e.g., from the claim\footnote{We find hard to believe that authors of the
OPERA experiment could be so naive spending so many million dollars \ in an
experiment and yet did not take into account the relativistic effects on the
proper time of the clocks used in their experiment. But, if the did such an
elementary mistake what could we think of the care in the preparation of high
energy experiments?} in\ \cite{contaldi} that their authors do not take into
account the subtleties related to the problem of synchronization of clocks in
a non inertial reference frame \cite{rodsha2001} to the claim that
\ superluminal neutrinos induces a \emph{bremsstrahllung} process
($\nu\rightarrow\nu+e^{-}+e^{+}$) that according to Cohen and Glashow
\cite{coh} allows one to exclude what they called the OPERA anomaly and place
a strong constraint on neutrino superluminality. This claim needs a more
careful analysis. Indeed, if superluminal particles exists, to make
predictions involving them we need first to build a consistent field theory
including a preferred reference frame \cite{maltrod1997,rodcap2007}, as, e.g.,
in \cite{cab,cab2006,ciboremb1996,ciboremb1998,ciboremb2011}. Indeed, the
authors of these last references show that with a preferred frame it is
possible to have not only a theory which includes tachyons free of causal
paradoxes but also free of the vacuum instability problem.

Returning to the subject of our paper, in Section 3 we recall how to obtain
easily subluminal and superluminal solutions as \textit{free boundary
solution}s for the homogeneous wave equation and then use those results to
build analogous solutions for the Weyl equation. In Section 4 we even present
a superluminal solution for Weyl equation with a front moving at superluminal
speed. In Section 5 we show how the Clifford bundle formalism used to describe
Weyl equation naturally leads to the conclusion that local chirality
invariance of that equation implies that the particles which it describes must
\ carry a magnetic charge which couples to the electromagnetic field.
Moreover, Weyl fields which are eigenstates of the parity operator describe a
pair of monopole anti-monopole system with null total magnetic charge. This
result may eventually be important to understand the propagation of neutrinos
in the interior of stars.

\section{Mathematical Preliminaries}

To start, we suppose that all phenomena occurs in a manifold $M\simeq
\mathbb{R}^{4}$. Let $\{x^{\mu}\}$ be global coordinate functions for $M$ and
let $\{\boldsymbol{e}_{\mu}=\partial/\partial x^{\mu}\}$ be a global basis for
$TM$ and $\{\gamma^{\mu}=dx^{\mu}\}$ a basis for $T^{\ast}M$ dual to the basis
$\{\boldsymbol{e}_{\mu}\}$. We equip $M$ with a Lorentz metric field
$\boldsymbol{g\in}\sec T_{0}^{2}M$ such that $\boldsymbol{e}_{\mu}%
\cdot\boldsymbol{e}_{\nu}:=\boldsymbol{g}(\boldsymbol{e}_{\mu},\boldsymbol{e}%
_{\nu})=\eta_{\mu\nu}$\ and with a field $g\in\sec T_{2}^{0}M$ \ such that
$\gamma^{\mu}\cdot\gamma^{\nu}:=g(\gamma^{\mu},\gamma^{\nu})=\eta^{\mu\nu}$
where the matrices with entries\ $\eta_{\mu\nu}$ and $\eta^{\mu\nu}$ are
diagonal matrices denoted by $\mathrm{diag}(1,-1,-1,-1)$. We introduce also
the bases $\{\boldsymbol{e}^{\mu}\}$ of $TM$ and $\{\gamma_{\mu}\}$ of
$T^{\ast}M$ that are respectively reciprocal to the bases $\{\boldsymbol{e}%
_{\mu}\}$ and $\{\gamma^{\mu}\}$, i.e., $\boldsymbol{e}^{\mu}\cdot
\boldsymbol{e}_{\nu}=\delta_{\nu}^{\mu}$ and $\gamma^{\mu}\cdot\gamma_{\nu
}=\delta_{\nu}^{\mu}$. Of course the manifold $M$ is oriented by the volume
element $\gamma^{5}:=\tau_{g}=\gamma^{0}\gamma^{1}\gamma^{2}\gamma^{3}\in\sec%
{\textstyle\bigwedge\nolimits^{4}}
T^{\ast}M$ and also time oriented ($\uparrow$) by $\partial/\partial x^{0}$
and we recognize the pentuple $\langle M,\boldsymbol{g},D,\tau_{g}%
,\uparrow\rangle$, where $D$ is the Levi-Civita connection of $\boldsymbol{g}$
as the Minkowski spacetime structure.

In what follows we suppose that all fields involved in our calculations are
sections or equivalence classes of sections of the Clifford bundle of
differential forms\footnote{Eventually for some computations it is a good idea
(as is the case in electrodynamic) to use complex functions as the components
of the Clifford fields. This corresponds to work in $\mathbb{C\otimes
}\mathcal{C\ell}(M,g)$.} $\mathcal{C}\ell(M,g)$. In section 4 we use for easy
of calculations the complexified Clifford bundle $\mathbb{C\otimes}%
\mathcal{C}\ell(M,g)$.

Covariant Dirac spinor fields, i.e., the Dirac fields used in books of field
theories in the Clifford bundle formalism are represented by Dirac-Hestenes
spinor fields (\emph{DHSF}) and covariant Weyl spinor fields are represented
by \emph{DHSF} satisfying an algebraic constraint to be specified below. A
\emph{DHSF} on Minkowski spacetime is an equivalence class of pairs $(\Xi
_{u},\psi_{_{\Xi_{u}}})$, where $\Xi_{u}$ is a spinorial frame\footnote{The
symbol $u$ in $\Xi_{u}$ denotes an element of \textrm{Spin}$_{1,3}^{e}$. The
space $\mathbf{\Theta}$ of spinorial frames can be thought as an
\textit{extension} of the space $\mathcal{B}$ of orthonormal \textit{vector
frames }of $T^{\ast}M$\textit{ }, where even if two vector frames have the
\textit{same} ordered vectors, they are considered distinct if the spatial
axes of one vector frame is rotated by an odd number of $2\pi$ rotations
relative to the other vector frame and are considered the same if the spatial
axes of one vector frame is rotated by an even number of $2\pi$ rotations
relative to the other frame. Even if this construction seems to be impossible
at first sight, Aharonov and Susskind \cite{as} warrants that it can be
implemented physically. More details may be found in
\cite{r2004,mr2004,rodcap2007}.} field and $\psi_{_{\Xi_{u}}}$ is an
appropriate sum of even nonhomogeneous multiform fields,
\begin{equation}
\psi_{_{\Xi_{u}}}=S+F+\gamma^{5}P \label{1}%
\end{equation}
where $S,P\in\sec%
{\textstyle\bigwedge\nolimits^{0}}
T^{\ast}M\hookrightarrow\sec\mathcal{C}\ell(M,g)$ and $F\in\sec%
{\textstyle\bigwedge\nolimits^{2}}
T^{\ast}M\hookrightarrow\sec\mathcal{C}\ell(M,g)$. Let $\{\Gamma^{\mu}\}$ be
an arbitrary orthonormal coframe for Minkowski spacetime (denoted fiducial
coframe) to which we associate a spin frame\ $\Xi_{u_{0}}$ where $u_{0}=1$ is
the identity element of \textrm{Spin}$_{1,3}^{e}$. Let $\{\gamma^{\mu}\}$
\ and $\{\gamma^{\prime\mu}\}$ be two others orthonormal coframes for
Minkowski spacetime with associated spin frames $\Xi_{u}$ and $\Xi_{u^{\prime
}}$ such that%
\begin{equation}
\gamma^{\prime\mu}=(\pm u)\gamma^{\mu}(\pm u^{-1})=\Lambda_{\nu}^{\mu}%
\gamma^{\nu}, \label{2}%
\end{equation}
and where the matrix with entries $\Lambda_{\nu}^{\mu}$ is an element of
$\mathfrak{L}_{+}^{\uparrow}$, the homogeneous orthochronous Lorentz group.
Then, $(\Xi_{u},\psi_{_{\Xi_{u}}})\approx(\Xi_{u^{\prime}},\psi_{_{\Xi
_{u^{\prime}}}})$, if and only if,%
\begin{equation}
\psi_{_{\Xi_{u^{\prime}}}}u^{\prime-1}=\psi_{_{\Xi_{u}}}u^{-1}. \label{3}%
\end{equation}
In what follows we choose $\{\gamma^{\mu}\}$ as the fiducial coframe and put
$\psi_{_{\Xi_{u}}}=\psi$. As it is now well-known Dirac equation in our
formalism is represented by the so-called Dirac-Hestenes equation which reads:%
\begin{equation}
\boldsymbol{\partial}\psi\gamma^{2}\gamma^{1}-m\psi\gamma^{0}=0, \label{4}%
\end{equation}
where $m$ is the rest mass of the Dirac particle and
\begin{equation}
\boldsymbol{\partial}:=\gamma^{\mu}D_{\boldsymbol{e}_{\mu}}, \label{5}%
\end{equation}
is the Dirac operator acting on section of $\sec\mathcal{C}\ell(M,g)$.
Moreover, for an arbitrary Clifford field $\mathcal{C}\in\sec\mathcal{C}%
\ell(M,g)$ we have
\begin{equation}
\boldsymbol{\partial}\mathcal{C}=\boldsymbol{\partial}\wedge\mathcal{C+}%
\boldsymbol{\partial}\lrcorner\mathcal{C}=d\mathcal{C}-\delta\mathcal{C},
\label{7}%
\end{equation}
where $d=\boldsymbol{\partial}\wedge\mathcal{\ }$is the differential operator
and $\delta=-\boldsymbol{\partial}\lrcorner$ is the Hodge codifferential.

Now, the algebraic constraint defining a Weyl spinor field (denoted by
$\mathcal{F}_{\pm}\mathcal{\in}\sec(%
{\textstyle\bigwedge\nolimits^{0}}
T^{\ast}M+%
{\textstyle\bigwedge\nolimits^{2}}
T^{\ast}M+%
{\textstyle\bigwedge\nolimits^{4}}
T^{\ast}M)\hookrightarrow\sec\mathcal{C}\ell(M,g)$) from a given Dirac spinor
field $\psi$\ is:%
\begin{equation}
\mathcal{F}_{\pm}=\frac{1}{2}(\psi\mp\gamma_{5}\psi\gamma_{21}), \label{8}%
\end{equation}
and they are \textquotedblleft eigenvectors\textquotedblright\ of the
chirality operator $\gamma_{5}$, i.e.,
\begin{equation}
\gamma_{5}\mathcal{F}_{\pm}=\pm\mathcal{F}_{\pm}\gamma_{21}. \label{9}%
\end{equation}
Moreover, observe that Weyl spinor fields satisfy the important
relation\footnote{The symbol \textquotedblleft$\widetilde{}$\textquotedblright%
\ denotes the reverse operator. If $A_{r}\in\sec%
{\textstyle\bigwedge\nolimits^{r}}
T^{\ast}M\hookrightarrow\sec\mathcal{C\ell}(M,g)$ we have that $\tilde{A}%
_{r}=(-1)^{\frac{r(r-1)}{2}}A_{r}$.}%
\begin{equation}
\mathcal{\tilde{F}}_{\pm}\mathcal{F}_{\pm}=\mathcal{F}_{\pm}\mathcal{\tilde
{F}}_{\pm}=0. \label{10}%
\end{equation}

We recall moreover that in the Clifford bundle formalism the parity operator
$\mathbf{P}$ \cite{lounesto} is represented in such a way that for
Dirac-Hestenes spinor field $\psi$ just defined above we have%

\begin{equation}
\mathbf{P}\psi(t,\mathbf{x)}=-\gamma_{0}\psi(t,-\mathbf{x)}\gamma_{0}\;.
\label{parity1}%
\end{equation}

The following Dirac-Hestenes spinor fields are eigenstates of the parity
operator with eigenvalues $\pm1$:%

\begin{equation}%
\begin{array}
[c]{c}%
\mathbf{P}\psi^{\uparrow}=+\psi^{\uparrow}\;,\quad\psi^{\uparrow}=\gamma
_{0}\psi_{-}\gamma_{0}-\psi_{-}\;,\\
\mathbf{P}\psi^{\downarrow}=-\psi^{\downarrow}\;,\quad\psi^{\downarrow}%
=\gamma_{0}\psi_{+}\gamma_{0}+\psi_{+}\;.
\end{array}
\label{parity2}%
\end{equation}

Having saying that the Weyl equation describing a massless neutrino
$(\mathcal{F=F}_{+})$ field is
\begin{equation}
\boldsymbol{\partial}\mathcal{F}=0. \label{11}%
\end{equation}

Given, $A,B\in\sec%
{\textstyle\bigwedge\nolimits^{1}}
T^{\ast}M\hookrightarrow\sec\mathcal{C}\ell(M,g)$ we define a generalized
potential \cite{rv1997} for the Dirac-Hestenes spinor field,
\begin{equation}
\mathcal{A}=(A+\gamma_{5}B)\mathcal{\in}\sec(%
{\textstyle\bigwedge\nolimits^{1}}
T^{\ast}M+%
{\textstyle\bigwedge\nolimits^{3}}
T^{\ast}M)\hookrightarrow\sec\mathcal{C}\ell(M,g), \label{12}%
\end{equation}
such that
\begin{equation}
\psi=\boldsymbol{\partial}\mathcal{A} \label{13}%
\end{equation}
For the Weyl field $\mathcal{F}_{\pm}$ we have
\begin{equation}
\mathcal{F}_{\pm}= \boldsymbol{\partial}\mathcal{A}_{\pm}^{\prime}\label{13a}%
\end{equation}
where $\mathcal{A}_{\pm}^{\prime}= A_{\pm}^{\prime}+ \gamma_{5} B_{\pm
}^{\prime}$ with
\begin{equation}
2A^{\prime}_{\pm}= A \pm A\cdot\gamma_{03}\pm B\cdot\gamma_{21} ,
\quad2B^{\prime}_{\pm}= B \mp A\cdot\gamma_{21} \pm B\gamma_{03} .
\end{equation}

Since $\boldsymbol{\partial}\mathcal{F}_{\pm}=0$ we have that%
\begin{equation}
\square A_{\pm}^{\prime}=0\text{, \ }\square B_{\pm}^{\prime}=0. \label{14}%
\end{equation}

\subsection{Energy-Momentum Tensor of the Weyl Field}

Now, the adjunct\footnote{The adjunct of an extensor field $E:%
{\textstyle\bigwedge\nolimits^{1}}
T^{\ast}M\rightarrow\sec%
{\textstyle\bigwedge\nolimits^{1}}
T^{\ast}M$ is the extensor field $E^{\dagger}:\sec%
{\textstyle\bigwedge\nolimits^{1}}
T^{\ast}M\rightarrow\sec%
{\textstyle\bigwedge\nolimits^{1}}
T^{\ast}M$ such that for all $\boldsymbol{n,m}\in\sec%
{\textstyle\bigwedge\nolimits^{1}}
T^{\ast}M$ we have: $E^{\dagger}(\boldsymbol{n})\cdot\boldsymbol{m}%
=\boldsymbol{n}\cdot E(\boldsymbol{m})$.} of the energy-momentum $1$-form
fields $T_{\mu}^{\dagger}$ $\mathcal{\in}\sec(%
{\textstyle\bigwedge\nolimits^{1}}
T^{\ast}M+%
{\textstyle\bigwedge\nolimits^{3}}
T^{\ast}M)\hookrightarrow\sec\mathcal{C}\ell(M,g)$ of the Weyl field is given
by (see, e.g.,\ Chapter 7 of \cite{rodcap2007}):%
\begin{equation}
T^{\dagger}(\gamma_{\mu})=T_{\mu}^{\dagger}=\langle\partial_{\mu
}\mathcal{F\gamma}_{2}\gamma_{1}\gamma_{0}\mathcal{\tilde{F}}\rangle_{1},
\label{e1}%
\end{equation}
and it not symmetrical (in general). The density of energy of the Weyl field
is then given by
\begin{equation}
T_{0}^{\dagger}\cdot\gamma_{0}=T_{00} \label{e2}%
\end{equation}
and the energy of a Weyl field configuration $\mathcal{F}$ is%
\begin{equation}
\mathcal{E}=%
{\textstyle\int}
T_{00}dx^{1}dx^{2}dx^{3}. \label{e3}%
\end{equation}

\section{Superluminal Solutions of the Weyl Equation}

It is now well known \cite{marod1996,rodcap2007} that all relativistic wave
equations have \textit{boundary free} solutions with arbitrary speeds $0\leq
v<\infty$. The set of such solutions has three disjoint classes, the
subluminal, luminal and superluminal ones. Each solution\ within one of the
classes may be transformed in other one within the same class to which it
belongs by the action of the Lorentz group. However, not all solutions within
the superluminal class can be realized in nature according to the Principle of
Relativity as a physical phenomenon in an arbitrary inertial frame, for \ if
one of those solutions represent the description of some \emph{real}
phenomenon carrying information it would be possible to send information to
the past (see a discussion of that issue in \cite{rodcap2007}). Thus, if all
Lorentz deformed solutions are realized as physical phenomena in any given
inertial reference frame then we must arrive at the conclusion that we have a
breakdown of the Principle of Relativity and identification of a
\emph{preferred }inertial frame in our universe which gives the natural time
order of events. In the preferred frame all Lorentz deformed solutions always
correspond to possible phenomena.\medskip

In this section we recall two very simple superluminal solutions of the scalar
wave equation, which may be used almost immediately (as we shall see) to build
superluminal solutions for Weyl equation. The energy of the simple solutions
as calculated using Eq.(\ref{e3}) is infinite, as it is the case of the energy
of all plane wave solutions of all relativistic wave equations. However, we
think that a \emph{quantum mechanic} interpretation for that extraordinary
solutions may be given associating the energy and momentum of a giving wave
carrying one neutrino through the dispersion relation of the solution using
the well-known formulas, $\mathcal{E=\hbar\omega}$ and $\left\vert \vec
{p}\right\vert =\mathcal{\hbar}k$.

\subsection{Subluminal and Superluminal Spherical Bessel Beams}

Consider the homogeneous wave equation (HWE)
\begin{equation}
{\frac{\partial^{2}}{\partial t^{2}}}\Phi-\nabla^{2}\Phi=0\;. \label{A2}%
\end{equation}
We now present some subluminal and superluminal solutions of Eq.(\ref{A2})
called \textit{subluminal and superluminal spherical Bessel
Beams\footnote{Historical details about the discovery of these solutions and
other non referenced statements below may be found in \cite{rl1997}.}.} To
introduce these beams we define the variables
\begin{equation}
\xi_{<}=[x^{2}+y^{2}+\gamma_{<}^{2}(z-v_{<}t)^{2}]^{1/2}\;; \label{A29a}%
\end{equation}%
\begin{equation}
\gamma_{<}={\frac{1}{\sqrt{1-v_{<}^{2}}}}\;;\ \ \omega_{<}^{2}-k_{<}%
^{2}=\Omega_{<}^{2}\;;\ \ v_{<}={\frac{d\omega_{<}}{dk_{<}}}\;; \label{A29b}%
\end{equation}%
\begin{equation}
\xi_{>}=[-x^{2}-y^{2}+\gamma_{>}^{2}(z-v_{>}t)^{2}]^{1/2}\;; \label{A29c}%
\end{equation}%
\begin{equation}
\gamma_{>}={\frac{1}{\sqrt{v_{>}^{2}-1}}}\;;\ \ \omega_{>}^{2}-k_{>}%
^{2}=-\Omega_{>}^{2}\;;\ \ v_{>}=d\omega_{>}/dk_{>}\;. \label{A29d}%
\end{equation}

We can now easily verify that the functions $\Phi_{<}^{\ell_{m}}$ and
$\Phi_{>}^{\ell_{m}}$ below are respectively subluminal and superluminal
solutions of the HWE (see how to obtain these solutions, e.g., in
\cite{rl1997}). We have
\begin{equation}
\Phi_{p}^{\ell m}(t,\vec{x})=C_{\ell}\,j_{\ell}(\Omega_{p}\xi_{p}%
)\,P_{m}^{\ell}(\cos\theta)e^{im\theta}e^{i(\omega_{p}t-k_{p}z)} \label{A30}%
\end{equation}
where the index $p=<$, or $>$, $C_{\ell}$ are constants, $j_{\ell}$ are the
spherical Bessel functions, $P_{m}^{\ell}$ are the Legendre functions and
$(r,\theta,\varphi)$ are the usual spherical coordinates. $\Phi_{<}^{\ell m}$
$[\Phi_{>}^{\ell m}]$ has phase velocity $(w_{<}/k_{<})<1$ $[(w_{>}/k_{>})>1]$
and the modulation function $j_{\ell}(\Omega_{<}\xi_{<})$ $[j_{\ell}%
(\Omega_{>}\xi_{>})]$ moves with group velocity $v_{<}$ $[v_{>}]$, where
$0\leq v_{<}<1$ $[1<v_{>}<\infty]$. Both $\Phi_{<}^{\ell m}$ and $\Phi
_{>}^{\ell m}$ have been called \textit{undistorted progressive waves} (UPWs).
This term has been introduced by Courant and Hilbert; however they didn't
suspect of UPWs moving with speeds greater than $c=1$. For the simple
applications that we have in mind we shall need the form of $\Phi_{<}^{00}$
and $\Phi_{>}^{00}$, which we denote simply by $\Phi_{<}$ and $\Phi_{>}$:%

\begin{equation}
\Phi_{p}(t,\vec{x})=C\frac{\sin(\Omega_{p}\xi_{p})}{\xi_{p}}e^{i(\omega
_{p}t-k_{p}z)};\ \ p=<\;\mbox{or}\;>\;. \label{A31}%
\end{equation}
When $v_{<}=0$, we have $\Phi_{<}\rightarrow\Phi_{0}$, with%

\begin{equation}
\Phi_{0}(t,\vec{x})=C{\frac{\sin\Omega_{<}r}{r}}e^{i\Omega_{<}t}%
,\;r=(x^{2}+y^{2}+z^{2})^{1/2}\;. \label{A32}%
\end{equation}
When $v_{>}=\infty$, $\omega_{>}=0$ and $\Phi_{>}^{0}\rightarrow\Phi_{\infty}%
$, with%

\begin{equation}
\Phi_{\infty}(t,\vec{x})=C_{\infty}{\frac{\sinh\rho}{\rho}}e^{i\Omega_{>}%
z},\ \ \rho=(x^{2}+y^{2})^{1/2}\;. \label{A33}%
\end{equation}

\begin{remark}
We observe that if our interpretation of phase and group velocities is
correct, then there must be a Lorentz frame where $\Phi_{<}$ is at rest. It is
trivial to verify that in the coordinate chart $\langle x^{\prime\mu}\rangle$
which is a natural adapted coordinate chart to the inertial reference frame
$\boldsymbol{e}_{0}^{\prime}=\partial/\partial t^{\prime}$,%
\[
\partial/\partial t^{\prime}=(1-v_{<}^{2})^{-1/2}\partial/\partial
t+(v_{<}/\sqrt{1-v_{<}^{2}})\partial/\partial z
\]
which is an inertial Lorentz frame moving with speed $v_{<}$ in the $z$
direction relative to $\boldsymbol{e}_{0}=\partial/\partial t$, $\Phi_{p}$
goes into $\Phi_{0}(t^{\prime},\vec{x}^{\prime})$ given by \emph{Eq.(\ref{A32}%
)} with $t\mapsto t^{\prime}$, $\vec{x}\mapsto\vec{x}^{\prime}$. \newline
\end{remark}

\subsection{Subluminal and Superluminal Bessel Beams}

The solutions that are necessary for the developments of the next section are
solutions of the HWE, in cylindrical coordinates. Here we briefly recall how
these solutions are obtained in order to present subluminal and more
important, for what concern the objectives of this paper, superluminal
solutions of Weyl equation.\ In what follows the cylindrical coordinate
functions are denoted as usual by $(\rho,\theta,z)$, $\rho=(x^{2}+y^{2}%
)^{1/2}$, $x=\rho\cos\theta$, $y=\rho\sin\theta$. We write for $\Phi$:
\begin{equation}
\Phi(t,\rho,\theta,z)=f_{1}(\rho)f_{2}(\theta)f_{3}(t,z)\;. \label{A34}%
\end{equation}
Inserting \ Eq.(\ref{A34}) in Eq.(\ref{A2}) gives%

\begin{subequations}
\label{A35}%
\begin{gather}
\rho^{2}{\frac{d^{2}}{d\rho^{2}}}f_{1}+\rho{\frac{d}{d\rho}}f_{1}+(L\rho
^{2}-\nu^{2})f_{1}=0,\label{a}\\
\left(  {\frac{d^{2}}{d\theta^{2}}}+\nu^{2}\right)  f_{2}=0,\label{b}\\
\left(  {\frac{d^{2}}{dt^{2}}}-{\frac{\partial^{2}}{\partial z^{2}}}+L\right)
f_{3}=0. \label{c}%
\end{gather}
\ 

In these equations $L$ and $\nu$ are separation constants. Since we want
$\Phi$ to be periodic in $\theta$ we choose $\nu=n$ an integer. For $L$ we
consider two cases:

\subsubsection{Subluminal Bessel Solution, $L=\Omega_{<}^{2}>0$}

In this case Eq.(\ref{A35}) is a Bessel equation and we have
\end{subequations}
\begin{equation}
\Phi_{J_{n}}^{<}(t,\rho,\theta,z)=C_{n}J_{n}(\rho\Omega_{<})e^{i(k_{<}%
z-w_{<}t+n\theta)}\text{ }n\in\mathbb{N}, \label{A36}%
\end{equation}
where $n\in\mathbb{N}$ and $C_{n}$ is a constant, $J_{n}$ is the $n$-th order
Bessel function and
\begin{equation}
\omega_{<}^{2}-k_{<}^{2}=\Omega_{<}^{2}\;. \label{A37}%
\end{equation}
The $\Phi_{J_{n}}^{<}$ are eventually called in acoustical papers the
$nth$-order non-diffracting Bessel beams.

Bessel beams are examples of UPWs. They are subluminal waves. Indeed, the
group velocity for each wave is
\begin{equation}
v_{<}=d\omega_{<}/dk_{<}\,,\ 0<v_{<}<1\;, \label{A38}%
\end{equation}
but the phase velocity of the wave is $(\omega_{<}/k_{<})>1$. That this
interpretation is correct follows from an argument similar to the one just
presented above for the case of the spherical beams.\medskip

It is convenient for what follows to define the variable $\eta$, called the
axicon angle \cite{rl1997},
\begin{equation}
k_{<}=\overline{k}_{<}\cos\eta\;,\ \ \Omega_{<}=\overline{k}_{<}\sin
\eta\;,\ \ 0<\eta<\pi/2\;. \label{A39}%
\end{equation}
Then
\begin{equation}
\overline{k}_{<}=\omega_{<}>0 \label{A40}%
\end{equation}
and Eq.(\ref{A36}) can be rewritten as $\Phi_{A_{n}}^{<}\equiv\Phi_{J_{n}}%
^{<}$, with
\begin{equation}
\Phi_{A_{n}}^{<}=C_{n}J_{n}(\overline{k}_{<}\rho\sin\eta)e^{i(\overline{k}%
_{<}z\cos\eta-\omega_{<}t+n\theta)}. \label{A41}%
\end{equation}
In this form the solution is sometimes known in acoustical papers as the
$n$-th order non-diffracting portion of the \textit{axicon Beam}. The phase
velocity $v^{ph}=1/\cos\eta$ is independent of $\overline{k}_{<}$, but, of
course, it is dependent on $k_{<}$. We shall show in Section 4 that
surprisingly as it may be. waves constructed from appropriated superpositions
$\Phi_{J_{n}}^{<}$ beams may be \textit{superluminal} !

\subsubsection{Superluminal (Modified) Bessel Solution, $B=-\Omega_{>}^{2}<0$}

In this case Eq.(\ref{A35}) is the modified Bessel equation and we denote the
solutions by
\begin{equation}
\Psi_{K_{n}}^{>}(t,\rho,\theta,z)=C_{n}K_{n}(\Omega_{>}\rho)e^{i(k_{>}%
z-\omega_{>}t+n\theta)}, \label{A42}%
\end{equation}
with $n\in\mathbb{N}$ and where $K_{n}$ are the modified Bessel functions,
$C_{n}$ are constants and
\begin{equation}
\omega_{>}^{2}-k_{>}^{2}=-\Omega_{>}^{2}\;. \label{A43}%
\end{equation}
We see that $\Phi_{K_{n}}^{>}$ are also examples of UPWs, each of which has
group velocity $v_{>}=d\omega_{>}/dk_{>}$ such that $1<v_{>}<\infty$ and phase
velocity $0<(\omega_{>}/k_{>})<1$. As in the case of the spherical Bessel beam
[Eq.(\ref{A31})] we see again that our interpretation of phase and group
velocities is correct. Indeed, for the superluminal (modified) Bessel beam
there is no inertial Lorentz frame where the wave is stationary. The solution
$\Psi_{K_{0}}^{>}$ will be denoted simply by $\Psi^{>}$ in what follows.

\subsection{A Weyl Superluminal Solution}

Let us choose $A^{\prime}$ and $B^{\prime}$ such that $A_{0} = A_{3} = B_{0} =
B_{3} = 0$. Moreover, let us choose $A_{2} = \mp B_{1}$ or $A_{1} = \pm B_{2}%
$. In this case we can write
\begin{equation}
A_{\pm}^{\prime}= \Upsilon m, \quad B^{\prime}_{\pm}= \Upsilon n ,\label{A440}%
\end{equation}
where we have omitted the subscript $\pm$ on $\Upsilon_{\pm}$ and $m$ and $n$
are constant 1-form fields such that $m \cdot n = 0$.  Then Eq.(\ref{14})
implies that%
\begin{equation}
\square\Upsilon=0\text{.} \label{A45}%
\end{equation}

If we choose for $\Upsilon$, e.g., either $\Phi^{>}$ or $\Psi^{>}$ we
immediately have that the corresponding Weyl field
\begin{equation}
\mathcal{F}_{\pm}=\boldsymbol{\partial}\Upsilon_{\pm}(m+\gamma_{5}n)
\label{A46}%
\end{equation}
is propagating at superluminal speed.

\begin{remark}
What is really interesting in these solutions is that the dispersion relation
simulates the propagation of a \emph{(}tachyonic\emph{) }massive neutrino. A
fitting of the parameters for the two solutions that are compatible with the
OPERA experiment and analogous ones may help to clarify the issue of
incompatible square neutrino masses seem in different situations.
\end{remark}

\begin{remark}
It is important to recall that for the case of the Dirac-Hestenes equation,
for non singular solutions, i.e., the ones such that $\Psi_{D}\tilde{\Psi}%
_{D}\neq0$, we can write $\Psi_{D}=\rho^{1/2}\exp(\beta\gamma/2)R$, where
$\rho$ and $\beta$ are scalar functions and $\forall x\in M$, $R(x)\in
\mathrm{Spin}_{1,3}^{e}$. As a consequence the current is $J_{D}=\Psi
\gamma^{0}\tilde{\Psi}=\rho v$, and the vector field $v=R\gamma^{0}\tilde{R}$
is obviously always timelike independently of $\Psi_{D}$ being a subluminal,
luminal or superluminal solution of the Dirac-Hestenes equation. Thus in the
de Broglie-Bohm interpretation of quantum mechanics electrons even if the have
associated a superluminal wave travels at subluminal speed. For the case of a
Weyl spinor field the current $J_{W}=\mathcal{F}\gamma^{0}\mathcal{\tilde{F}}$
is always lightlike in view of \emph{Eq.(\ref{10})} independently of
$\mathcal{F}$ being a subluminal, luminal or superluminal solution of the Weyl
equation. Such observations may lead one to believe that superluminal
phenomena associated to neutrino propagation may be a kind of reshaping
phenomenon like the ones that happens in electron or photon tunneling
experiments \emph{\cite{cr1998,rtx2001,crtx2001}}.
\end{remark}

\section{Solutions with Front Moving at Superluminal Speed}

In this section we present superluminal solutions of Weyl equation where the
fronts of the waves move at superluminal speeds. To exhibit one such
solution,\ in order to simplify the calculations we complexify Weyl equation,
i.e., we take$\ $the components of $\mathcal{F}$ as sections of
$\mathbb{C\otimes}\mathcal{C}\ell(M,g)$. Next we choose the potentials $A$ and
$B$ as in the last section as%
\begin{equation}
A^{\prime}_{\pm}= \Psi m, \quad B^{\prime}_{\pm}= \Psi n \label{guess}%
\end{equation}
where $m$ and $n$ are constant $1$-form fields such that $m \cdot n = 0 $ and
$\Psi$ is a solution of%
\begin{equation}
\square\Psi=0\text{,} \label{WE}%
\end{equation}
but differently of the last section where we found boundary free solutions of
the HWE (and thus boundary free solutions of Weyl equation), here we want to
look for a solution of Eq.(\ref{WE}) which solves a \emph{Sommerfeld like
problem}, i.e., one satisfying at the $z=0$ plane the following boundary
conditions (given in cylindrical coordinates):
\begin{equation}%
\begin{array}
[c]{l}%
{\footnotesize \Psi(t,\rho,0)=T(t)}\int\limits_{-\infty}^{\infty
}{\footnotesize d\omega D(\omega)J}_{0}{\footnotesize (\omega\rho}%
\sin{\footnotesize \eta)e}^{-i\omega t}{\footnotesize ,}\\
\left.  \frac{\partial\Psi(t,\rho,z)}{\partial z}\right\vert _{z=0}\\
{\footnotesize =iT(t)}\cos{\footnotesize \eta}\int\limits_{-\infty}^{\infty
}{\footnotesize d\omega D(\omega)J}_{0}{\footnotesize (\omega\rho}%
\sin{\footnotesize \eta)k(\omega)e}^{-i\omega t}{\footnotesize ,}%
\end{array}
\label{swe1}%
\end{equation}
where $\mathbf{T}(t)=\left[  \Theta(t+T)-\Theta(t-T)\right]  $ , $\Theta$ is
the Heaviside function, $k(\omega)=\omega$, and $\eta$ is a constant called
the axicon angle \cite{caprod2001} and $D(k)$ is an appropriate frequency
distribution to be determined in order for $\mathcal{E}$ to result finite. As
showed in \cite{crtx2001} a complex solution of Eq.(\ref{WE}) (for $z>0,t>T$)
which satisfies the Sommerfeld conditions (Eqs.(\ref{swe1})) is in polar
coordinates given by%

\begin{gather}
{\small \Psi(t,\rho,z)=}%
{\textstyle\int\nolimits_{-\infty}^{\infty}}
{\small d\omega D(\omega)J}_{0}{\small (\omega\rho}\sin{\small \eta
)e}^{-i\omega(t-z\cos\eta)}{\small ,}\nonumber\\
\text{ for }\left\vert t-z\cos\eta\right\vert {\small <T,}\label{solution}\\
{\small \Psi(t,\rho,z)=0,}\text{ for }\left\vert t-z\cos\eta\right\vert
{\small >T.}\nonumber
\end{gather}
We call $\Psi$ a superluminal $X$-pulse, its wave front obviously propagates
with superluminal speed in the $z$-direction.

A neutrino in this model has speed $v=1/\cos\eta$ and the axicon angle may be
found in order to fit the OPERA or analogous experiments.

\section{Chiral Invariance and the Interaction of Neutrinos with the
Electromagnetic Field.}

In this section we would like to call attention that eventually the
propagation mode of neutrinos in Earth (and more generally in regions
containing a electromagnetic field) are being wrongly evaluated. Such
statement will become clear with the theory just presented below which
suggests that Weyl equation describes a pair of opposite magnetic charged
sub-particles coupled together. Even if that effect is not appreciable for
neutrinos travelling from Geneve to the Gran Sasso, it may be eventually
important for neutrinos travelling in the medium of stars, and thus its
consequences seems worth to be investigated, something that we defer to
another publication.

Since \ $\boldsymbol{\partial}\mathcal{F}=0$ we see that Weyl equation is
invariant under constant duality transformations $\mathcal{F\mapsto F}%
^{\prime}=\exp(\vartheta\gamma_{5})\mathcal{F}$ \ where $\vartheta$ is a
constant. Moreover taking into account that $\mathcal{F}$ is eigenvector of
the chirality operator (Eq.(\ref{parity2})) we can write Weyl equation as%
\begin{equation}
\boldsymbol{\partial}\mathcal{F\gamma}_{21}=\boldsymbol{\partial
}\mathcal{\gamma}_{5}\mathcal{F}=0. \label{w1}%
\end{equation}
Now, if we make a spacetime time dependent duality transformation
$\mathcal{F\mapsto F}^{\prime}=\exp(\beta\gamma_{5})\mathcal{F}$ where $\beta$
is a smooth function on $M$, we see that Weyl equation will be invariant only
if we introduce a compensating potential field $\mathbf{B}\in\sec
\bigwedge\nolimits^{1}T^{\ast}M\hookrightarrow\sec\mathcal{C}\ell
(M,\boldsymbol{g})$, i.e., we need to have the equation
\begin{equation}
\boldsymbol{\partial}\mathcal{F}_{+}+g\mathcal{\gamma}_{5}\mathbf{B}%
\mathcal{F}_{+}=0. \label{sw8.6}%
\end{equation}

This equation as it is easily verified is invariant under the gauge
transformations
\begin{equation}
\mathcal{F}_{+}\mapsto\mathcal{F}_{+}e^{g\gamma_{5}\theta};\text{
\ }\mathbf{B}\mapsto\text{ }\mathbf{B}+\partial\theta. \label{sw8.7}%
\end{equation}

Also, the equation for $\mathcal{F}_{-}$ coupled with an electromagnetic
potential $\mathbf{B}\in\sec\bigwedge\nolimits^{1}T^{\ast}M\hookrightarrow
\sec\mathcal{C}\ell(M,\boldsymbol{g})$ is
\begin{equation}
\boldsymbol{\partial}\mathcal{F}_{-}-g\mathcal{\gamma}_{5}\mathbf{B}%
\mathcal{F}_{-}=0. \label{sw8.8}%
\end{equation}
which is invariant under the gauge transformations
\begin{equation}
\mathcal{F}_{-}\mapsto\mathcal{F}_{-}e^{g\gamma_{5}\theta};B\mapsto
B-\partial\theta. \label{sw8.9}%
\end{equation}
showing clearly that the fields $\mathcal{F}_{+}$ and $\mathcal{F}_{-}$ carry
\textit{opposite} `charges'. Consider now the Weyl field in interaction with
the potential $\mathbf{B}$ and where the spinor fields $\mathcal{F}^{\uparrow
},\mathcal{F}^{\downarrow}$ (recall Eq.(\ref{parity2})) are eigenvectors of
the \emph{parity} operator (recall Eq.(\ref{parity1})) and look for solutions
of Eq.(\ref{sw8.6}) such that $\mathcal{F}=\mathcal{F}^{\uparrow}$, i.e., the
equation
\begin{equation}
\boldsymbol{\partial}\mathcal{F}^{\uparrow}\gamma_{21}+g\mathbf{B}%
\mathcal{F}^{\uparrow}=0. \label{sw8.10}%
\end{equation}
This equation separates in two equations,
\begin{equation}
\boldsymbol{\partial}\mathcal{F}_{+}^{\uparrow}+g\gamma_{5}\mathbf{B}%
\mathcal{F}_{+}^{\uparrow}=0;\text{ \ \ \ }\boldsymbol{\partial}%
\mathcal{F}_{-}^{\uparrow}-g\gamma_{5}\mathbf{B}\mathcal{F}_{-}^{\uparrow}=0,
\label{sw8.11}%
\end{equation}
showing that a Weyl spinor field that is an eigenvector of the parity operator
describes a \textit{pair} of particles with opposite `charges'. We interpret
these particles (following Lochak\footnote{Lochak suggested that an equation
equivalent to Eq.(\ref{sw8.10}) describe massless monopoles of opposite
`charges'.} \cite{lochak}) as \emph{massless} `monopoles' of opposite magnetic
charges interacting\ with an electromagnetic field.

After this discoursing we may suggest that when propagating from Geneve to the
Gran Sasso the neutrinos are interacting with the Earth's electromagnetic
field which may cause some effect on their propagation (eventually tunneling
of its wave packet). Even if that effect be very small in that case it may be
important when neutrinos travel in the strong electromagnetic field of stars.
So, our suggestion seems in principle worth to be investigated, something we
defer to another publication.

\section{Conclusions}

The superluminal solutions that we exhibit above may be\ eventually useful for
describing a neutrino flying from Geneve to the Gran Sasso in the Opera
experiment \cite{adam} supposing the data is indeed reliable. In our model it
has speed $v=1/\cos\eta$ and the axicon angle may be discovered in order to
fit data of the experiment. It goes without saying that the eventual existence
of finite energy superluminal solutions of Weyl equations implies in a
breakdown of the Principle of Relativity thus selecting one preferred
reference frame $\mathbf{P}\in\sec TM$ ($\mathbf{P}\cdot\mathbf{P}=1$). We may
conjecture that this preferred frame is to be identified with the timelike
component $g(\Gamma^{0},$ ) of the fiducial vector coframe $\{\Gamma^{\mu}\}$
associated with the spinorial frame $\Xi_{u_{0}}$ mentioned in the
introduction. However before claiming that we indeed observed a breakdown of
Lorentz invariance it is necessary to investigate more carefully if one can
find an explanation consistent with the Principle of Relativity, i.e., if we
can explain the results of the OPERA experiment as a kind of pulse reshaping
where, as in the electromagnetic case (recall footnote 3), the group velocity
may be superluminal (but where the front velocity of the neutrino wave is
always the usual light velocity), or even if the result of that experiment is
simply due to a superluminal group velocity resulting from \emph{neutrino
oscillations}, something that can be obtained, e.g., by making the potentials
$A_{\pm}$ and $B_{\pm}$ in Eq.(\ref{14}) to satisfy wave equations with
dispersion relations with different masses\footnote{See also in this respect
\cite{meco}.}. Finally we recall that in the last section it was shown that
Weyl equation may be thought as describing the propagation of a pair of
opposite magnetic charge\footnote{Thus, eventually carrying a magnetic dipole
moment \cite{eidel}.} neutrinos that may interact with an external
electromagnetic field. Thus the electromagnetic field of the Earth may have
some influence on the propagation of neutrinos, and although the effect may be
negligible for the case of Earth's electromagnetic field it may be eventually
worth to investigate that interaction for the case of neutrinos emitted from
stars, which are submitted to very strong electromagnetic fields.

At least, we mention that once the preferred frame is identified, it is a good
idea for observers in all inertial frames use as time coordinate the time
coordinate defined in the preferred frame. With the use of this time there are
no causality paradox, all signals propagates only to the future as defined by
the preferred time. We also mention that the transformations relating the
spacetime coordinates of two different inertial frames moving relative to each
other (and relative to the preferred frame) realize a nonstandard realization
of the Lorentz group. This issue has been discussed in
\cite{maltrod1997,rodcap2007}.


\begin{thebibliography}{99}                                                                                               %


\bibitem {adam}{\footnotesize Adam, T., et al, \textit{Measurement of the
Neutrino Velocity with the OPERA Detector in the CNGS Beam}.
[arXiv:1109.4897v1[hep-ex]]}

\bibitem {as}{\footnotesize Aharonov, Y., and Susskind, L., Observability of
the Sign of Spinors under a }${\footnotesize 2\pi}$ {\footnotesize Rotation,
\textit{Phys. Rev.} \textbf{158}, 1237-1238 (1967).}

\bibitem {assa}{\footnotesize Assamagan, K. A., et al, Upper Limit of the
Muon-Neutrino Mass and Charged-Pion Mass from Momentum Analysis of a Surface
Muon Beam,\textit{ Phys. Rev. D}, 6065-6077 (1966).}

\bibitem {bionta}{\footnotesize Bionta, R., Blewitt, G., Bratton, C., Capser,
D., Ciocio A., et al, Observation of a Neutrino Burst in Coincidence with
Supernova SN 1987a in the Large Magellanic Cloud, \textit{Phys. Rev. Lett.}
\textbf{58}, 1494-1496 (1987).}

\bibitem {cab}{\footnotesize Caban, P., Rembieli\'{n}ski, J., and
Smoli\'{n}ski, \textit{Decays of Spacelike Neutrino}%
s.[arXiv:hep-ph/97077391v1]}

\bibitem {cab2006}{\footnotesize Caban, P., Rembieli\'{n}ski, and
Smoli\'{n}ski, K.A, Oscillations do not Distinguish Between Massive and
Tachyonic Neutrinos,\textit{ Found. Phys. Lett}. \textbf{19,} 619-623 (2006).}

\bibitem {ciboremb1996}{\footnotesize Ciborowski, J., and Rembieli\'{n}ski,
J., \textit{Experimental Results and the Hypothesis of Tachyonic Neutrinos,
talk presented at ICHEP} 96. [arXiv:hep-ph/9607477v1]}

\bibitem {ciboremb1998}{\footnotesize Tritium Decay and the Hypotheis of
Tachyonic Neutrinos, \textit{Eur. Phys .J. C }\textbf{8}, 157-161 (1999).
[arXiv:hep-ph/9810355v1]}

\bibitem {ciboremb2011}{\footnotesize Ciborowski, J., and Rembieli\'{n}ski,
J.,\textit{ Comments on the Recent Velocity Measurement of the Muon Neutrinos
by the OPERA Collaborations}. [arXiv:1109.5599v1 [hep-ex]]}

\bibitem {coh}{\footnotesize Cohen, A. G. Glashow, S. L., \textit{New
Constraints on Neutrino Velocities}. [arXiv:1109.6562v1 [hep-ph]]}

\bibitem {contaldi}{\footnotesize Contaldi, C. R., \textit{The OPERA Neutrino
Velocity Result and the Synchronization of Clocks}. [arXiv:1109.6160v1
[hep-ph]]}

\bibitem {eidel}{\footnotesize Eidelman, S. et al. (Particle Data Group)
(2004). Review of Particle Physics, 2004-2005,\textit{ Phys. Lett. B}
\textbf{592}, 1--5.(2004)}

\bibitem {enders1993}{\footnotesize Enders, A., and Nimtz, G., Photonic
Tunneling Experiments, \textit{Phys. Rev. B} \textbf{47}, 9605-9609 (1993).}

\bibitem {hira}{\footnotesize Hirata, K., et al (KAMIOKANDE-II Collaboration)
, Observation of a Neutrino Burst from Supernova SN 1987, \textit{Phys. Rev.
Lett.} \textbf{58}, 1490-1493 (1987).}

\bibitem {longo}{\footnotesize Longo, M. J., Tests of Relativity from SN1987a,
\textit{Phys. Rev. D} \textbf{36}, 3276-3277 (1987).}

\bibitem {iorio}{\footnotesize Iorio, L., \textit{Environmental Fifth-Force
Hypothesis for the OPERA Superluminal Neutrino Phenomenology: Constraints from
Orbital Motions Around the Earth}.[arXiv:1109.6249 [gr-qc]]}

\bibitem {Keh}{\footnotesize Kehaglas, A., \textit{Relativistic Superluminal
Neutrino}. [arXiv:1109.6312v1[hep-ph]]}

\bibitem {kono}{\footnotesize Konoplya, R. A., \textit{Superluminal Neutrinos
and the Tachyon's Stability in the Rotating Universe.} [arXiv:1109.6215v1
[hep-ph]]}

\bibitem {lochak}{\footnotesize Lochak, G., Wave Equation for a Magnetic
Monopole, \textit{Int. J. Theor. Phys.} \textbf{24}, 1019-1050 (1985).}

\bibitem {lounesto}{\footnotesize Lounesto P., \textit{Clifford Algebras and
Spinors}, Cambridge Univ. Press, Cambridge, 1997.}

\bibitem {marod1996}{\footnotesize Maiorino, J. E., and Rodrigues, W. A. Jr.,
\ A Unified Theory for the Construction of Arbitrary Speeds }%
${\footnotesize 0\leq v<\infty}$ {\footnotesize of the Relativistic Wave
Equations, \emph{Random Oper. and Stoch. Equ}. \textbf{4}, 355-400 (1996).}

\bibitem {maltrod1997}{\footnotesize Matolcsi, T., and Rodrigues, W. A. Jr.,
The Geometry of Spacetime with Superluminal Phenomena, \textit{Alg. Groups and
Geom}. \textbf{14}, 1-16 (1997). [arXiv:physics/9710024]}

\bibitem {meco}{\footnotesize Mecozzi, A. and Bellini, M., \emph{Superluminal
Group Velocity of Neutrinos}. [arXiv:11101253v1 [hep-ph]]}

\bibitem {mr2004}{\footnotesize Mosna, R. A., and Rodrigues, W \ A. \ Jr., The
Bundles of Algebraic and Dirac-Hestenes Spinor Fields, \textit{J. Math. Phys}.
\textbf{45}, 2945-2966. (2004). [arXiv:math-ph/0212033v5]}

\bibitem {mug}{\footnotesize Mugnai, D., Ranfagni, A., and Ruggeri, R.,
Observation of Superluminal Behaviors in Wave Propagation, \textit{Phys. Rev.
Lett.} \textbf{80}, 4930-4833 (2000).}

\bibitem {oiko}{\footnotesize Oikonomou, V. K., The 2d Gross-Neveu Model for
Pseudovector Fermions and Tachyonic Mass Generation.
[arXiv:1109.6170v1[hep-ph]]}

\bibitem {cr1998}{\footnotesize de Oliveira, E. Capelas de, and Rodrigues, W.
A. Jr., Superluminal Electromagnetic Waves in Free Space, \textit{Ann. der
Physik} \textbf{7}, 654-659 (1998).}

\bibitem {crtx2001}{\footnotesize de Oliveira, E. Capelas de, Rodrigues, W. A.
Jr., Thober, D. S., and Xavier, A. L. Jr., Thoughtful Comments on `Bessel
Beams and Signal Propagation', \textit{Phys. Lett. A} \textbf{284}, 296-303
(2001).}

\bibitem {caprod2001}{\footnotesize de Oliveira, E. Capelas, and Rodrigues, W
\ A. Jr., Finite Energy Superluminal Solutions of Maxwell Equations,
\textit{Phys. Lett. A} \textbf{291,} 367-370 (2001).}

\bibitem {rv1997}{\footnotesize Rodrigues, W. A. Jr, and Vaz, J. Jr.,
Subluminal and Superluminal Solutions in Vacuum of the Maxwell Equations and
the Massless Dirac Equations, \textit{Adv. Appl. Clifford Algebras}
\textbf{7}(S), 458-466 (1997).}

\bibitem {rl1997}{\footnotesize Rodrigues, W. A. Jr., and Lu, J.-Y., On the
Existence of Undistorted Progressive Waves (UPWs) of Arbitrary Speeds
}${\footnotesize 0\leq v<\infty}$ {\footnotesize in Nature, \textit{Found.
Phys.} \textbf{27}, 435-503 (1997).}

\bibitem {rodsha2001}{\footnotesize Rodrigues, W. A. Jr., and Sharif, M.,
Rotating Frames in RT: Sagnac's Effect in SRT and other Related Issues,
\textit{Found. Phys.} \textbf{31}},{\footnotesize 1767-1784 (2001).}

\bibitem {rtx2001}{\footnotesize Rodrigues, W. A. Jr., Thober,D. S., and
Xavier, A. L. Jr., Causal Explanation of Superluminal behavior of Microwave
Propagation in Free space, \textit{Phys. Lett. A }\textbf{284}, 217-224
(2001).}

\bibitem {r2004}{\footnotesize Rodrigues, W. A. Jr, Algebraic and
Dirac-Hestenes Spinors and Spinor Fields, \textit{J. Math. Phys}. \textbf{45},
2908-2944 (2004). [arXiv:math-ph/0212030v6]}

\bibitem {rodcap2007}{\footnotesize Rodrigues, W \ A. Jr., and Oliveira, E.
Capelas, \textit{The Many Faces of Maxwell, Dirac and Einstein Equations. A
Clifford Bundle Approach}, Lecture Notes in Physics \textbf{722}%
},{\footnotesize Springer, Heidelberg, 2007.}

\bibitem {stei1993}{\footnotesize Steiberg, A. M., Kwiat, P. G., and Chiao, R.
Y., Measurement of the Single Photon Tunneling Time, \textit{Phys. Rev. Lett.}
\textbf{71}, 708-711 (1993).}

\bibitem {tamblav}{\footnotesize Tamburini, F., and Laveder, M.,
\textit{Apparent Lorentz Violation with Superluminal Majorana Neutrinos at
OPERA}? [arXiv:1109.5445V3 [hep-ph]]}
\end{thebibliography}
\end{document}